\documentclass[twocolumn,showkeys,amsmath,amssymb]{revtex4-1}
\usepackage{graphicx}
\usepackage{dcolumn}
\usepackage{color}
\usepackage[colorlinks=true, citecolor=blue, linkcolor=blue, urlcolor=blue, anchorcolor=blue]{hyperref}
\usepackage{bm}
\usepackage{tabularx}
\usepackage{physics}
\usepackage{amsmath}
\usepackage{txfonts}

\begin{document}

\title{Engineering giant magnetic anisotropy in single-molecule magnets by dimerizing heavy transition-metal atoms}

\author{Jiaxing Qu and Jun Hu}
\email[]{E-mail: jhu@suda.edu.cn}
\affiliation{College of Physics, Optoelectronics and Energy, Soochow University, Suzhou, Jiangsu 215006, China \\
 Jiangsu Key Laboratory of Thin Films, Soochow University, Suzhou, Jiangsu 215006, China.}


\begin{abstract}
Search for single-molecule magnets with large magnetic anisotropy energy (MAE) is essential for the development of molecular spintronics devices used at room temperature. Through systematic first-principles calculations, we found that an Os-Os or Ir-Ir dimer embedded in the (5,5$^\prime$-Br$_2$-Salophen) molecule gives rise to large MAE of 41.6 or 51.4 meV which is large enough to hold the spin orientation at room temperature. Analysis of electronic structures reveals that the top Os and Ir atoms are most responsible for the spin moments and large MAEs of the molecules.
\end{abstract}

\maketitle

Single-molecule magnets (SMMs) provide discrete molecular spin states that can be used as quantum bits of information for storage, sensing and computing, which has inspired extensive interest in the context of next-generation data storage and communication devices \cite{Verdaguer, Review-1}, opening avenues for developing multifunctional molecular spintronics \cite{Rocha, Review-2, Review-3}. Such ideas have been researched extensively, using SMMs as localized spin-carrying centers for storage and for realizing logic operations \cite{Leuenberger}. Typical SMMs consist of organic networks and transition-metal (TM) atoms embedded in the organic networks. The electronic and magnetic properties of SMMs can be conveniently tuned by selecting appropriate organic networks, TM atoms or substrates \cite{Miller, Sessoli-1, HeinrichBW}. However, the development of molecular spintronics devices for practical applications is obstructed by the low blocking temperatures ($T_B$) which denotes the threshold of temperature for withholding thermal fluctuations of spin orientations \cite{Review-2, Review-3, Leuenberger}. Fundamentally, $T_B$ scales with the magnetic anisotropy energy (MAE), and $T_B$ of typical SMMs with 3d TM atoms is less than 10 K (or equivalently, MAE ~ 1 meV) \cite{HeinrichBW, Sessoli-2, KernK, Sessoli-3, LongJR, Gambardella}. To hold the spin orientation of SMMs at room temperature for spintronics applications, it is crucial to find SMMs with MAE larger than 30 meV \cite{Hu2014}.

It is known that the MAE originates from the spin-orbit coupling (SOC) \cite{MAE1, MAE2}. Hence, the potential candidates of SMMs are necessary to contain 4d or 5d TM atoms that possess large SOC constants. Recently, DiLullo $et~al.$ synthesized Co-(5,5$^\prime$-Br$_2$-Salophen) molecule successfully on Au(111) surface and studied the intriguing molecular Kondo effect \cite{Salophen}. This molecule has planar atomic structure with a triangle shape [see the inset in Fig. 1(a)], so it is convenient to build molecular junctions with this molecule. The simple synthesis progress ensures the possibility to obtain same structure with 4d and 5d TM atoms. Therefore, it is interesting to investigate the magnetic properties of these TM-(5,5$^\prime$-Br$_2$-Salophen) molecules (denoted as TM@Sal) to find possible candidates with large MAE.

In this paper, we studied the electronic and magnetic properties of the (5,5$^\prime$-Br$_2$-Salophen) molecule with a TM atom or TM dimer embedded, based on first-principles calculations. Although the MAEs of the TM@Sal molecules are small, the dimerization of two TM atoms results in significant enhancement of the MAEs. An Os-Os or Ir-Ir dimer embedded in the (5,5$^\prime$-Br$_2$-Salophen) molecule possesses large MAE of 41.6 or 51.4 meV which is large enough to guarantee these molecules for applications in molecular spintronics devices at room temperature. Analysis of the electronic structures reveals that the large MAEs originate from specific SOC effect in top Os and Ir atoms.

First-principles calculations were carried out with the Vienna {\it ab-initio} simulation package \cite{VASP1, VASP2}. The interaction between valence electrons and ionic cores was described within the framework of the projector augmented wave (PAW) method \cite{PAW1,PAW2}. The spin-polarized generalized-gradient approximation (GGA) was used for the exchange-correlation potentials \cite{PBE}. The energy cutoff for the plane wave basis expansion was set to 400 eV. The atomic positions were fully relaxed using the conjugated gradient method until the force on each atom is smaller than 0.01 eV/{\AA}. The MAE is defined as: $MAE=E_\parallel-E_\perp$, where $E_\parallel$ and $E_\perp$ denote the energies of in-plane and perpendicular spin orientations wit respective to the molecular plane, respectively. We used the torque method to calculate the MAEs of the TM@Sal molecules \cite{Torque1, Torque2}.

\begin{figure}
\centering
\includegraphics[width=8cm]{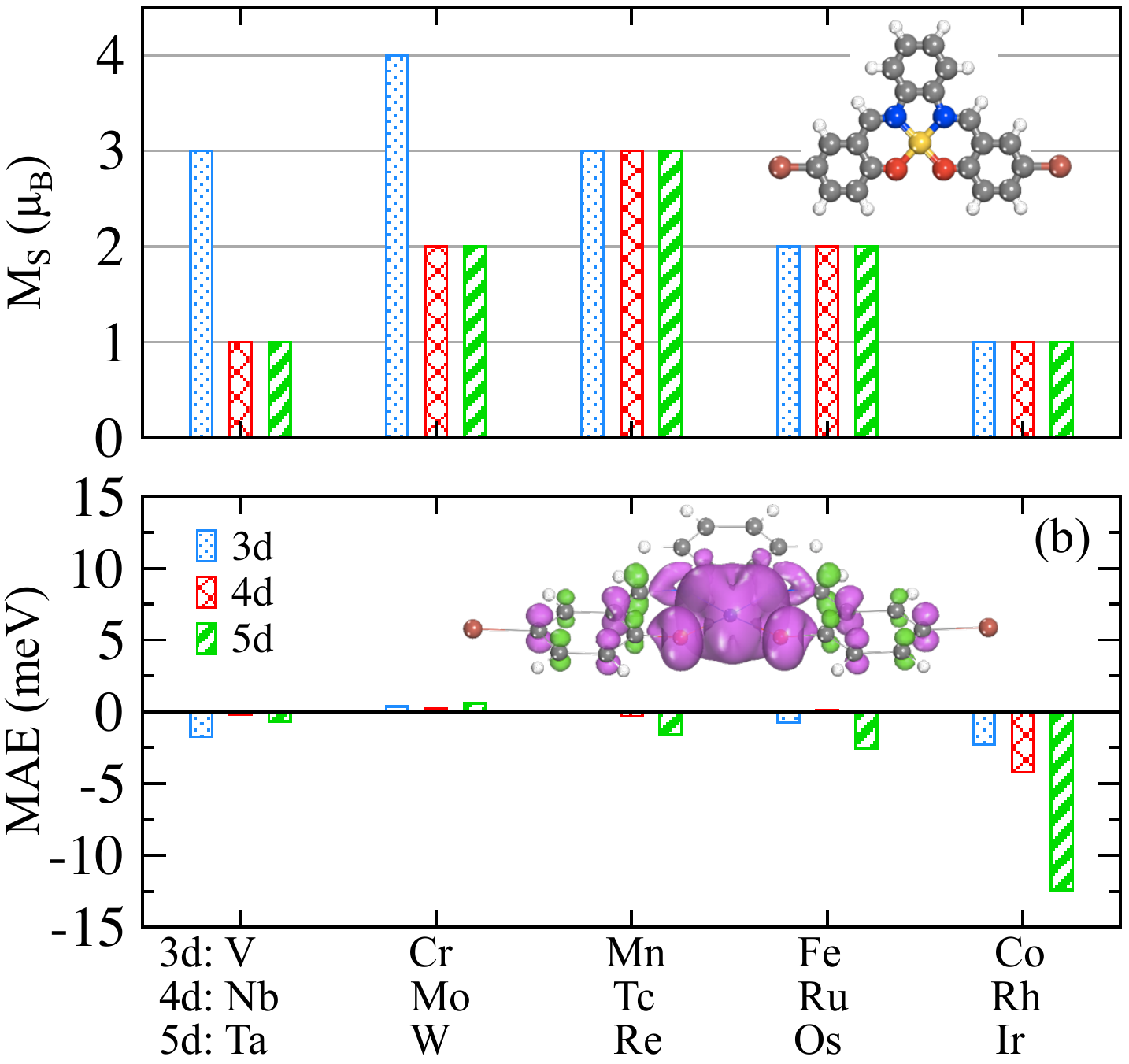}
\caption{(Color online) (a) Spin moment ($M_S$) and (b) magnetic anisotropy energy (MAE) of TM@Sal. The inset in (a) shows the top view of the atomic structure. The grey, blue, red, white, brown and golden spheres stand for C, N, O, H, Br and TM atoms, respectively. The inset in (b) shows the spin density of Os@Sal: $\rho_s=\rho_{\uparrow}-\rho_{\downarrow}$ with cutoff of $\pm0.01$ eV/$\AA^3$. The pink and green isosurfaces indicate the majority and minority spin densities, respectively.
}\label{mae1}
\end{figure}

We constructed a series of TM@Sal molecules and optimized the atomic structures. Due to the constraint of the organic network [see the inset in Fig. 1(a)], the TM-O and TM-N bond lengths vary little ($1.9\sim2.1$ {\AA}), even though the atomic radii are different much from 3d to 5d TM atoms. It can be seen from Fig. 1(a) that all the considered TM@Sal molecules are spin polarized. The largest spin moment ($M_S$) of 4 $\mu_B$ is on Cr@Sal. However, its 4d and 5d counterparts possess smaller $M_S$ of 2 $\mu_B$, mainly due to the smaller exchange field in 4d and 5d TM atoms than in 3d TM atoms. Similar situation exists in the V group, where the $M_S$ of V@Sal is 3  $\mu_B$ and that of Nb@Sal and Ta@Sal is 1 $\mu_B$. For the Mn, Fe and Co groups, the 4d and 5d TM atoms result in same $M_S$ as their 3d counterparts. Furthermore, we found that the spin moments of all TM@Sal molecules are mainly contributed from the TM atoms. Take Os@Sal for example, the Os atom in Os@Sal contributes about 1.6 $\mu_B$ to the total $M_S$, while the two O atoms also have visible contribution of about 0.3 $\mu_B$, as shown by the spin density in the set of Fig. 1(b).

To further explore the electronic and magnetic properties, we calculated the projected density of states (PDOS) of 5d orbitals in Os@Sal and Ir@Sal as plotted in Fig. 2. First of all, the $d_{xy}$ orbital are unoccupied and its energy level are far above the Fermi level ($E_F$), due to the strong repulsion along the TM-O and TM-N bonds. The $d_{yz}$ and $d_{z^2}$ orbitals are fully occupied, so they do not contributed to the total $M_S$. For Os@Sal, the $d_{xz}$ and $d_{x^2-y^2}$ orbitals are half occupied, with the majority spin states occupied and the minority spin states unoccupied, so each of them contributes 1 $\mu_B$ to the total $M_S$. Consequently, the total $M_S$ of Os@Sal is 2 $\mu_B$. For Ir@Sal, the minority spin state of $d_{x^2-y^2}$ is further occupied, because the Ir atom has one more electron than the Os atom. As a result, the total $M_S$ of Ir@Sal reduces to 1 $\mu_B$.

\begin{figure}
\centering
\includegraphics[width=8cm]{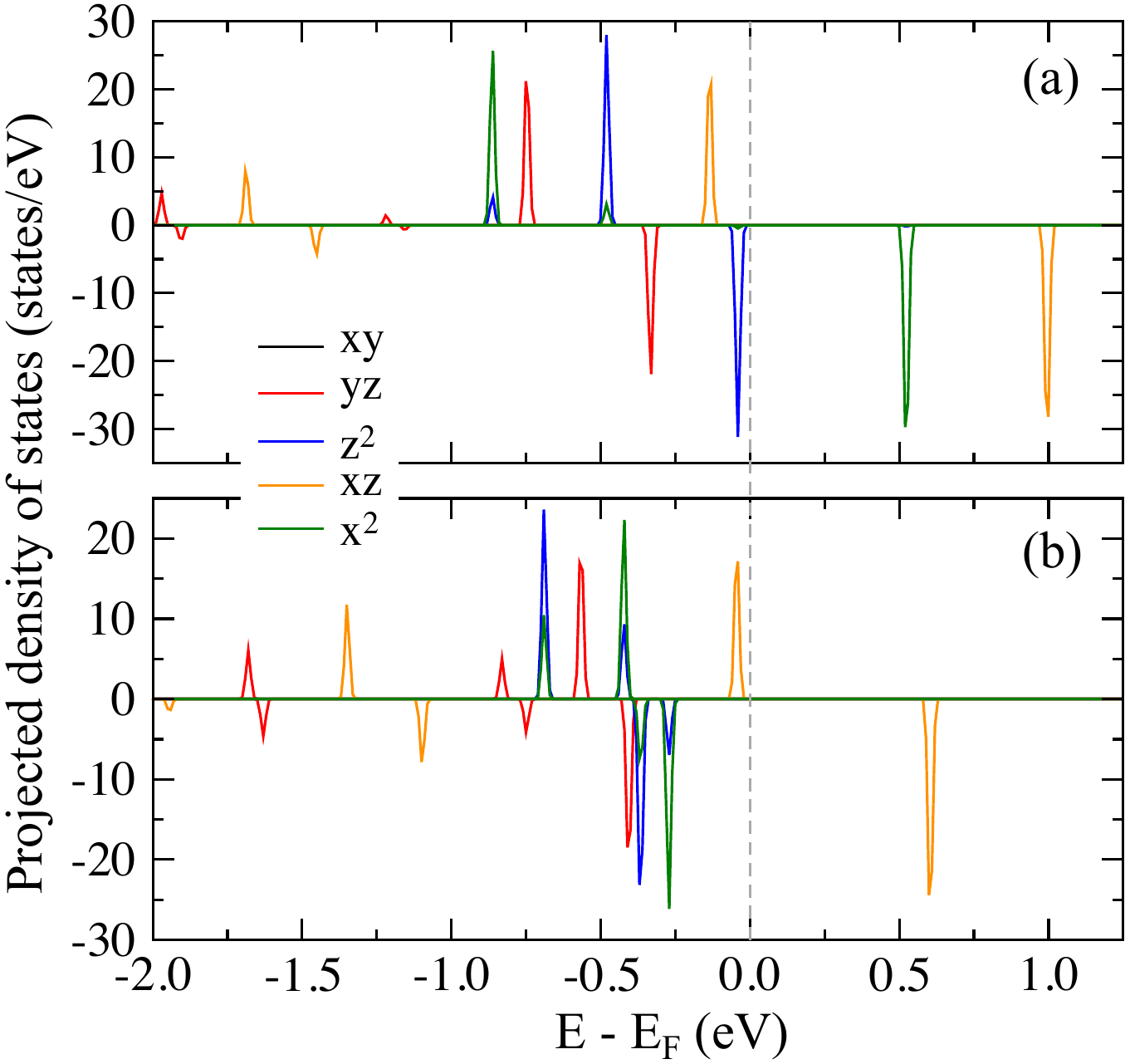}
\caption{(Color online) Projected density of states of the $5d$ orbitals in (a) Os@Sal and (b) Ir@Sal molecules, respectively. The vertical dashed line indicates the Fermi level ($E_F$).
}\label{dos1}
\end{figure}

Then we calculated the MAEs as plotted in Fig. 1(b). It can be seen that the Cr group of TM@Sal molecules have positive MAE, but the amplitude are quite small ($<$ 0.6 meV). The MAEs of the other molecules are negative. In the V group, V@Sal has larger MAE than the other two molecules. In the Mn, Fe and Co groups, the magnitudes of the MAEs of the 5d TM@Sal molecules are significantly larger than those of their 3d and 4d counterparts. In particular, the MAE of Ir@Sal is -12.4 meV, which implies the strong SOC effect in this molecule. However, large positive MAE is desired for practical applications in molecular spintronics devices, because the perpendicular spin orientation is easier to detect and control than the in-plane spin orientation.

The MAE originates from the SOC effect which has both perpendicular (positive) and in-plane (negative) contributions. The final MAE is the result of the competition between perpendicular and in-plane contributions, which can be expressed approximately in terms of matrix elements of angular momentum operators $L_x$ and $L_z$ based on the second-order perturbation theory \cite{Hu2014} 
\begin{equation}
MAE=\xi^2 \sum_{uo\alpha\beta} (-1)^{1-\delta_{\alpha\beta}} \bqty{\frac{\abs{\matrixel{u\alpha}{L_z}{o\beta}}^2- \abs{\matrixel{u\alpha}{L_x}{o\beta}}^2}{\varepsilon_{u\alpha}-\varepsilon_{o\beta}} }.
\end{equation}
Here $\xi$ is the SOC constant; $\varepsilon_{u\alpha}$ and $\varepsilon_{o\beta}$ are eigenvalues of unoccupied states with spin $\alpha$ ($\ket{u\alpha}$) and occupied states with spin $\beta$ ($\ket{o\beta}$, respectively. Obviously, the MAE can be further divided into three parts based on the spins ($\alpha$ and $\beta$) in the equation: (i) MAE(uu) contributed from the coupling between the majority spin states; (ii) MAE(dd) contributed from the coupling between the minority spin states; (iii) MAE(ud) contributed from the coupling between different spin channels. From the PDOS in Fig. 2, we found that the MAEs of both Os@Sal and Ir@Sal are dominated by the minority spin states, i.e. MAE(dd), while the MAE(uu) and MAE(ud) are negligible. This is because the majority spin orbitals are fully occupied, so that they do not contribute to the MAE. Further analysis reveals that the coupling between $d_{x^2-y^2}^u$ and $d_{yz}^o$ through $L_x$ operator, $\matrixel{d_{x^2-y^2}^u}{L_x}{d_{yz}^o}$, is responsible for the MAE(dd) of Os@Sal, while $\matrixel{d_{xz}^u}{L_y}{d_{z^2}^o}$ contributes most MAE(dd) of Ir@Sal.

\begin{figure}
\centering
\includegraphics[width=8cm]{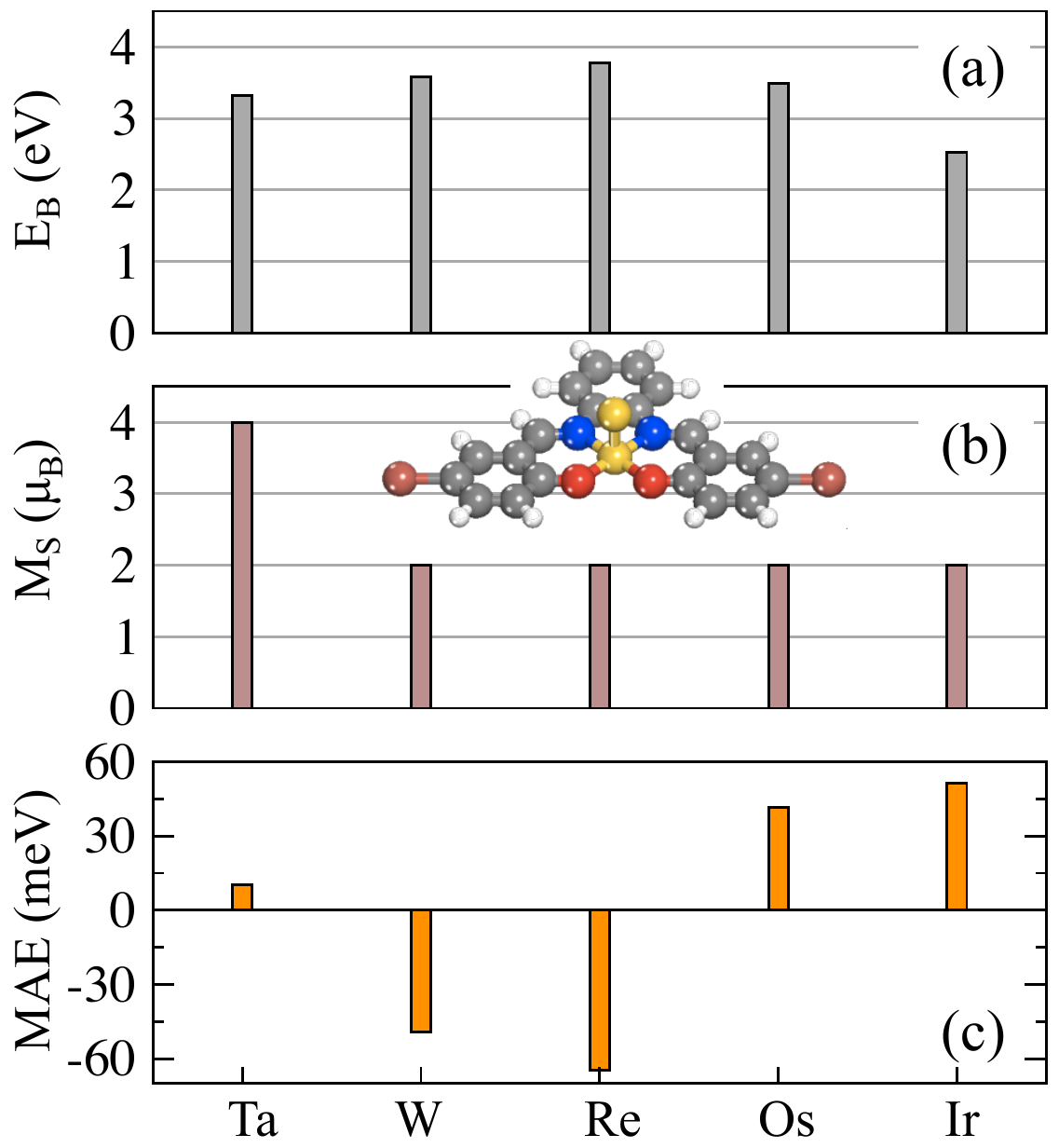}
\caption{(Color online) (a) Binding energy between the top TM atom and TM@Sal molecule. (b, c) Spin moment and magnetic anisotropy energy of TM$_2$@Sa. The inset shows the top view of the atomic structure. The grey, blue, red, white, brown and golden spheres stand for C, N, O, H, Br and transition-metal atoms, respectively.
}\label{mae2}
\end{figure}

In the growth progress of the TM@Sal molecules, it is possible for an extra TM atom to attach the molecule. The dimerization of the TM atoms changes the energy diagram of the molecule, hence changes the MAE as indicated in Eq. (1). Previous studies indeed revealed that the dimerization of two TM atoms results in large MAEs \cite{Hu2014, Hu2016}. Therefore, we studied the (5,5$^\prime$-Br$_2$-Salophen) molecules with 5d TM dimer embedded in the organic network, denoted as TM$_2$@Sal as displayed in Fig. 3. After relaxation of the atomic structures, the bottom TM atom is pulled out of the molecular plane by about 0.5 {\AA}. The TM-TM bond lengths vary from 2.1 to 2.5 {\AA}. To see the possibility of dimerization, we firstly calculated the bind energies as
\begin{equation}
E_b=E(TM)+E(TM@Sal)-E(TM_2@Sal),
\end{equation}
where E(TM), E(TM@Sal) and E(TM$_2$@Sal) are the total energies of free TM atom, TM@Sal molecule and TM$_2$@Sal molecule, respectively. As plotted in Fig. 3(a), the bind energies of all the 5d TM$_2$@Sal molecules are quite large, implying that the binding between the two TM atoms are very strong. So there is large chance to obtain TM$_2$@Sal molecules with standard synthesis progress \cite{Salophen}.

\begin{figure}
\centering
\includegraphics[width=8cm]{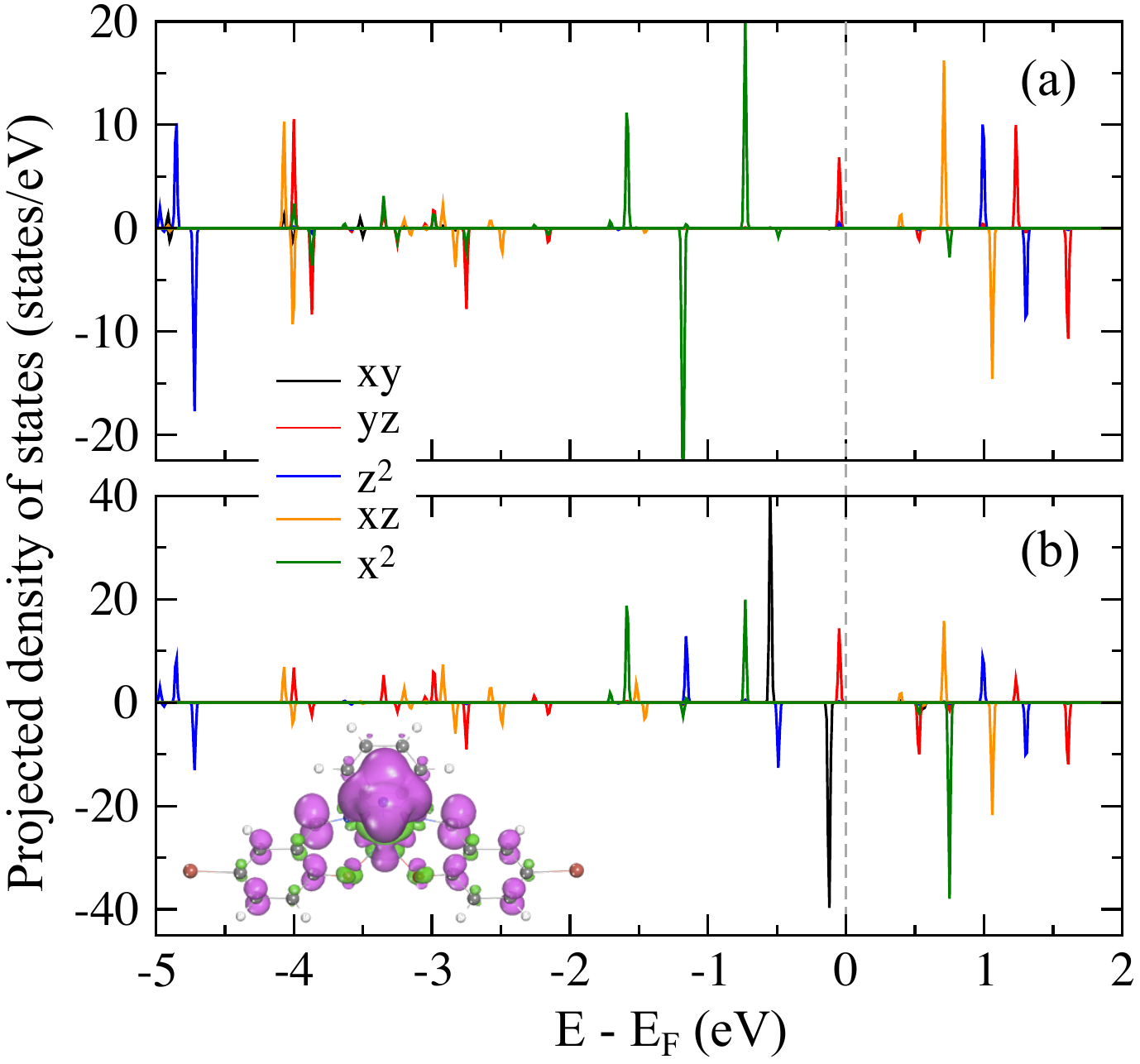}
\caption{(Color online) Projected density of states of the $5d$ orbitals in Os$_2$@Sal for the (a) bottom Os and (b) top Os atoms. The vertical dashed line indicates the Fermi level ($E_F$). The inset shows the spin density of Os$_2$@Sal with cutoff of $\pm0.01$ eV/$\AA^3$. The pink and green isosurfaces indicate the majority and minority spin densities, respectively.
}\label{dos2}
\end{figure}

Fig. 3(b) displays the total $M_S$ of the 5d TM$_2$@Sal molecules. The $M_S$ of the Ta$_2$@Sal molecule is 4 $\mu_B$, while those of all other 5d TM$_2$@Sal molecule are 2 $\mu_B$. However, the local spin moments of these molecules are different. The two TM atoms in W$_2$@Sal and Re$_2$Sal couple with each other antiferromagnetically, with local spin moments of $\sim 2.2$ and $\sim -0.2~\mu_B$ on the top and bottom TM atoms respectively. For Os$_2$@Sal, the bottom Os atom has small local spin moments of $\sim 0.05~\mu_B$, while the local spin moment is $\sim1.8~\mu_B$ on the top Os atom. In addition, some anion atoms have small contribution of $\sim 0.02~\mu_B$, which leads to a delocalized distribution of the spin density of Os$_2$@Sal as presented by the inset in Fig. 4. For Ir$_2$@Sal, on the other hand, the local spin moments on the top and bottom Ir atoms are $\sim 1.65$ and $\sim 0.30$, respectively, contributing almost the whole $M_S$ of the molecule, so the distribution of the spin density is quite localized as shown in the inset in Fig. 5.

The MAEs of the 5d TM$_2$@Sal molecules are shown in Fig. 3(c). Interestingly, the amplitudes of all the MAEs are large. The MAEs of Ta$_2$@Sal, Os$_2$@Sal and Ir$_2$@Sal are 10.3, 41.6 and 51.4 meV, respectively. The later two MAEs are large enough to hold the spin orientation perpendicular to the molecular plane at room temperature, so the Os$_2$@Sal and Ir$_2$@Sal molecules are promising for practical applications in molecular spintronics devices. Interestingly, the MAE of Ir$_2$@Sal is comparable to that of linear Ir-Ir and Ir-Cr nanowires \cite{Stepanyuk}. However, the linear Ir-Ir and Ir-Cr nanowires are more delicate to control, and large tensile strain is necessary for the linear Ir-Ir nanowire to possess large MAE. Therefore, SMMs such as Os$_2$@Sal and Ir$_2$@Sal are more promising for the sake of practical applications. On the other hand, the MAEs of W$_2$@Sal and Re$_2$@Sal are -49.3 and -64.7 meV, respectively. Clearly, the amplitudes of both MAEs are also large but with negative sign, indicating that the corresponding molecules prefer in-plane spin orientation. Therefore, the W$_2$@Sal and Re$_2$@Sal molecules are less favorable for applications.

\begin{figure}
\centering
\includegraphics[width=8cm]{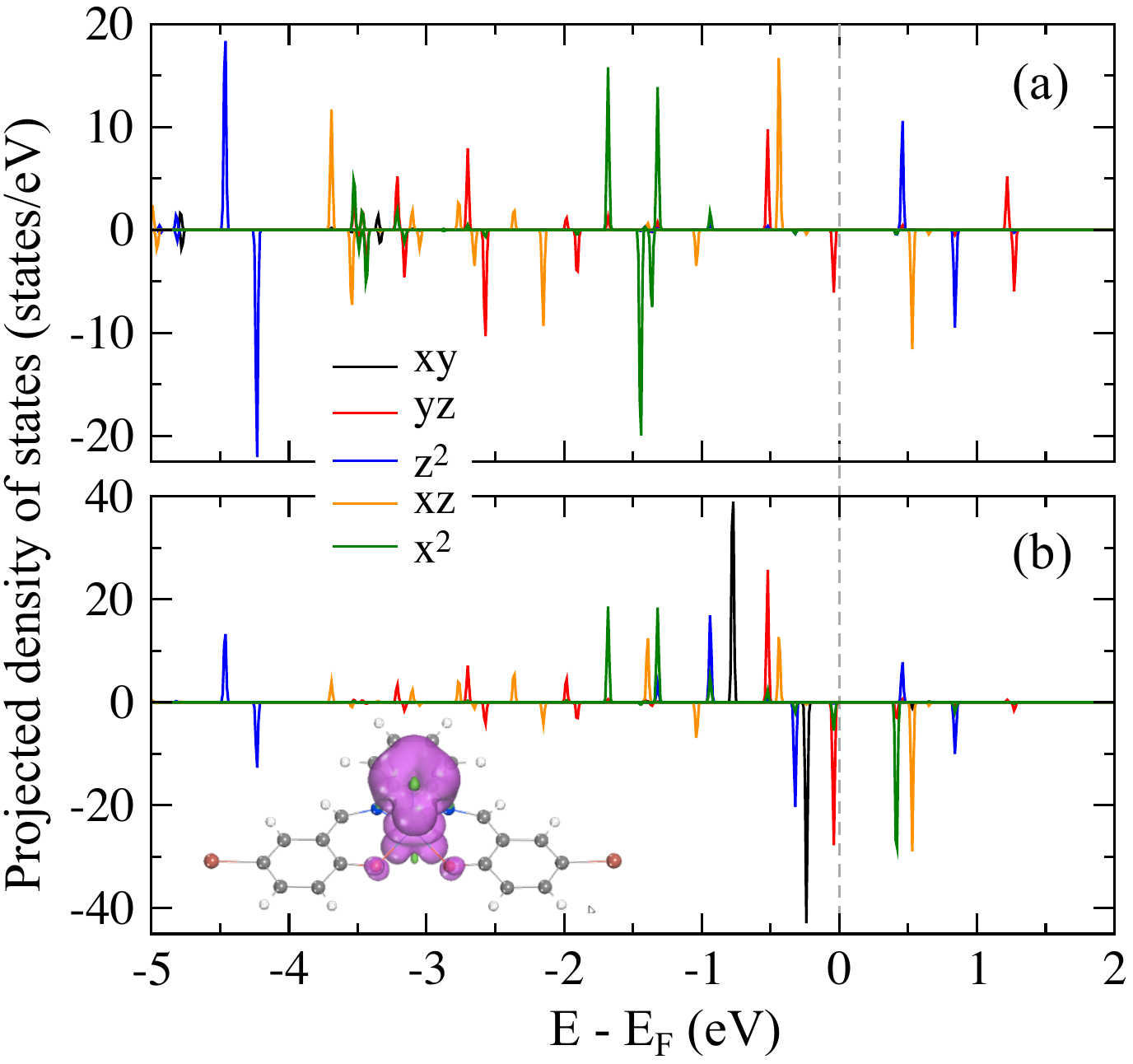}
\caption{(Color online)  Projected density of states of the $5d$ orbitals in Ir$_2$@Sal for the (a) bottom Ir and (b) top Ir atoms. The vertical dashed line indicates the Fermi level ($E_F$). The inset shows the spin density of Ir$_2$@Sal with cutoff of $\pm0.01$ eV/$\AA^3$. The pink and green isosurfaces indicate the majority and minority spin densities, respectively.
}\label{dos3}
\end{figure}

To reveal the origin of the giant positive MAEs in Os$_2$@Sal and Ir$_2$@Sal, we plotted their PDOS of the $5d$ orbitals of Os and Ir atoms in Fig. 4 and Fig. 5. Compared to Fig. 2, we can see that the PDOS of the bottom Os and Ir atoms are modified significantly, due to the strong interaction between the top and bottom TM atoms of the TM dimers. The strongest hybridization occurs between the $d_{z^2}$ orbitals of the top and bottom TM atoms, resulting in large energy separation between the bonding states and antibonding states ($>$ 5 eV). The hybridization between the $d_{xz/yz}$ orbitals are also strong. The interaction between the $d_{x^2-y^2}$ orbitals is relatively week, so that the energy separation is small ($\sim 1$ eV). The $d_{xy}$ orbital of the bottom TM atoms keeps unoccupied, due to the repulsion from the nearby N and O atoms. On the contrary, the $d_{xy}$ orbital of the top TM atoms is occupied in both majority and minority spin channels. Moreover, it can be seen from Fig. 4 and Fig. 5 that the total spin moments of these molecules are contributed from the $d_{yz}$ and $d_{x^2-y^2}$ orbitals and locates mainly on the top TM atoms. Therefore, the MAEs of these molecules are dominated by the top TM atoms. Based on Eq. (1), we identified that the coupling of $\matrixel{d_{x^2-y^2}^u}{L_z}{d_{xy}^o}$ of the top Os atom in the minority spin channel is most responsible for the giant positive MAE of Os$_2$@Sal. For Ir$_2$@Sal, the couplings of $\matrixel{d_{x^2-y^2}^u}{L_z}{d_{xy}^o}$ and $\matrixel{d_{xz}^u}{L_z}{d_{yz}^o}$ are the leading contribution to the MAE.

In summary, we studied the electronic and magnetic properties of the TM@Sal and TM$_2$@Sal molecules, based on first-principles calculations. It was found that the MAEs of the TM@Sal molecules are small and mostly negative. Interestingly, the dimerization of two TM atoms in a TM$_2$@Sal molecule results in significant enhancement of the MAE. In particular, the Os$_2$@Sal and Ir$_2$@Sal molecules possess giant MAEs of 41.6 and 51.4 meV, respectively. Analysis of the electronic structures reveals that both the spin moment and large MAE are dominated by the specific SOC effect in top Os and Ir atoms. Such large MAEs guarantee these molecules to be used in molecular spintronics devices at room temperature.

\section*{Acknowledgement}

This work is supported by the National Natural Science Foundation of China (11574223), the Natural Science Foundation of Jiangsu Province (BK20150303) and the Jiangsu Specially-Appointed Professor Program of Jiangsu Province. We also acknowledge the National Supercomputing Center in Shenzhen for providing the computing resources.


\end{document}